\documentclass[twoside,fleqn,a4paper]{article}

\usepackage{espcrc2MOD}
\usepackage{epsfig}
\usepackage{graphicx}
\usepackage{latexsym}

\begin{document}
\sloppy
\begin{titlepage}
\title{{\Large\bf Diffractive Structure Functions in the Model of the Stochastic Vacuum} \thanks{Talk presented at the QCD'98 Euroconference, Montpellier, July 1998.}}
\author{Oscar E. Ram\'{\i}rez del Prado \thanks{E-mail: ramirez@thphys.uni-heidelberg.de}$^{,}$ 
\thanks{supported by the Deutscher Akademischer Austauschdienst}$^{,}$ 
\address{Institut f\"ur Theoretische Physik der Universit\"at Heidelberg,\\ Philosophenweg 16, D-69120 Heidelberg}}
\begin{abstract}
We present a calculation for the diffractive structure functions $x_{\rm p} F_{\rm T}^{D(3)}$ and $x_{\rm p} F_{\rm L}^{D(3)}$  at low values of $Q^2$ using an eikonal approximation. The non-perturbative model of the stochastic QCD vacuum is used to evaluate the interactions between the quarks (antiquarks) and the color field. The diffractive final state contains only a quark-antiquark pair. We show the behavior of the diffractive structure functions against $\beta$ and $Q^2$.      
\end{abstract}
\end{titlepage}
\maketitle
\section{Introduction}
A calculation for the diffractive structure functions in a non-perturbative region using a leading eikonal approximation schema \cite{Nachtmann} and the model of the stochastic vacuum (MSV) \cite{Dosch1,DoschSimonov} is presented. The process under consideration is the diffractive dissociation of a photon by a proton. The proton is taken in a quark-diquark representation and at high energies the photon can be represented as a quark-antiquark pair. The quarks and the diquark pick up an eikonal phase factor in the vacuum background field and the interaction is obtained by average over the vacuum field by means of the MSV. If we draw the trajectories of the $q-\bar{q}$ pair in a space-time diagram and close the extremes to have a colorless object, we get the so called Wegner-Wilson Loops.

\section{Diffractive Structure Functions}

\begin{figure}[h]
\leavevmode 
\begin{center}
\includegraphics[width=6.5cm]{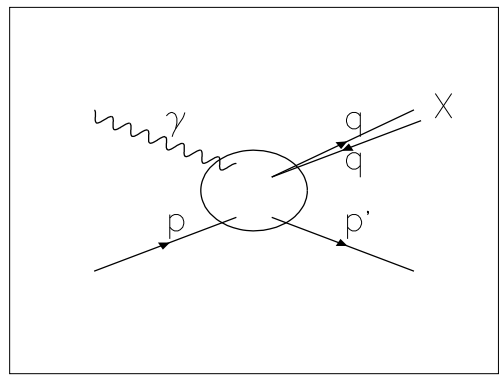}
\unitlength1cm
\vspace*{-2em}
{\caption{\scriptsize Diffractive dissociation process.} \label{fig1}}
\end{center}
\end{figure}

We want to calculate the diffractive disso\-cia\-tion process (figure \ref{fig1}) where a virtual photon scatters with a proton and in the final state we get a proton $+$ anything ($X$):

\begin{equation} \label{difdis}
\gamma^{*}+p\rightarrow p+X \; ,
\end{equation}

\noindent we describe the ``anything'' by a $q-\bar{q}$ pair in a color singlet state. In the center of mass frame the initial transverse momentum of both photon and proton is $\vec{0}$, furthermore for the photon $q^{+}$ and for the proton $p^{-}$ become large. After the collision the proton remains almost intact with momentum $p' \approx p$ and the photon dissociates in a quark carrying a fraction $z$ of the photon's original momentum and an antiquark carrying the remaining $(1-z)$ fraction of the momentum. This configuration is very sensitive to the extreme kinematical regions where the quark or the antiquark carries most of the photon momentum i.e, $z \rightarrow 1,0$ \cite{BuchmullerMcDemontHebecker}. 

\vspace*{-2em}
\vspace*{-2em} 
The momentum transfer is purely transversal, and it will be denoted by $\vec{\Delta}_{\perp}$.

We are working at ISR energies in $s$ and in the regime where $s\gg t$ with a momentum transfer $|t|\leq 1 \; GeV^2$.

We define the following invariants:

\begin{eqnarray} \label{invariants}
s&=&W^2=(p+q)^2 \; ,\\
t&=&(p'-p)^2=\Delta_{\perp}^2 \; ,
\end{eqnarray}

\noindent and the standard variables for diffraction:

\begin{equation} \label{xpomeron}
x_p=\frac{Q^2+M^2}{Q^2+W^2}=\frac{x}{\beta} \hspace*{1em} , \hspace*{1em} x=\frac{Q^2}{s} \; ,
\end{equation}

\begin{equation} \label{beta}
\beta=\frac{Q^2}{Q^2+M^2} \; ,
\end{equation}

\noindent where $M^2$ is the invariant mass of the $q-\bar{q}$ pair.

In a pomeron picture $x_p$ can be interpretated as the momentum fraction of the pomeron within the proton and $\beta$ as the momentum fraction of the struck quark within the pomeron.

The phase space is given by:

\begin{equation} \label{phasespacedef}
d\Phi_3=\frac{1}{2^3(2\pi)^5}\frac{Q^2}{2s}\frac{dx_p}{x_p}\frac{dz}{\beta}\;d^2\Delta_{\perp}\;d\theta_{\vec{l}} \; ,
\end{equation}

The cross section is 

\begin{equation} \label{crosssectiondef}
d\sigma^{\lambda}=\frac{1}{2s}\left|T^{\lambda}\right|^2d\Phi_3 \; ,
\end{equation}

\noindent and the scattering amplitude $T^{\lambda}$, following \cite{Nachtmann,DoschFerreraKramer}, is given by 

\begin{eqnarray} \label{tamplitude}
\lefteqn{T^{\lambda} = i2s(2\pi)^{\frac{1}{2}}\sqrt{2z(1-z)}\int{d^2b} \; e^{i{\vec{\Delta}}_{\perp}\cdot\vec{b}}} \nonumber \\
&&\times \int{d^2r_{\perp}}\Psi^{\lambda}_{\gamma^{*}}(z,r_{\perp})W(\vec{b},\vec{r_{\perp}},z)e^{-i\vec{l}\cdot\vec{r_{\perp}}} \; ,
\end{eqnarray}

\noindent the index $\lambda$ in the photon wave function $\Psi^{\lambda}_{\gamma^{*}}(z,r_{\perp})$ refers to its polarization ($\lambda=0$ longitudinal, $\lambda=\pm 1$ transversal).

The function 

\begin{equation} \label{functionw}
W(\vec{b},\vec{r_{\perp}},z) = \hspace*{-.5em} \int \hspace*{-.5em} \frac{d^2 r_2}{4\pi} \left|
\Psi_P(r_2)\right|^2 
\tilde{J}\left(\vec{r}_1,\vec{r}_2,\vec{b},z\right),
\end{equation}

\noindent involves the proton-dipole interaction and contains the loop-loop scattering term $\tilde{J}\left(\vec{r}_1,\vec{r}_2,\vec{b},z\right)$ which is evaluated in the MSV.

We take a Gaussian function for the proton quark-diquark wave function 

\begin{equation} \label{protonwf}
\Psi_P(r_2)=\frac{\sqrt{2}}{S_P}e^{-\frac{r_2^2}{4S_P^2}} \; ,
\end{equation}

\noindent and the photon wave function has the form given in \cite{DoschGousKulPir} according to light-cone perturba\-tion theory.

Now, writing everything together and integra\-ting over $\vec{\Delta}_{\perp}$ and $b$ we have 

\begin{eqnarray} \label{crossection} 
\lefteqn{d\sigma^{\lambda}=\frac{Q^2}{2^2(2\pi)^2}\frac{dx_p}{x_p}\frac{dz}{\beta}d\theta_{\vec{p}} \; z(1-z)} \\ 
&&\hspace*{-2em} \times \int{d^2b}\left| \int d^2r_{\perp} \Psi^{\lambda}_{\gamma^{*}}(z,r_{\perp})W(\vec{b},\vec{r_{\perp}},z)e^{-i\vec{l}\cdot\vec{r_{\perp}}}\right|^2 \; . \nonumber 
\end{eqnarray}

With the usual definition of the structure functions and taking into account the symmetries in the angular integrations and the different po\-la\-ri\-za\-tions we obtain:

\begin{description}
\item [a)] Longitudinal case:
\begin{eqnarray} \label{longitudinal}
\lefteqn{x_P F_{\rm L}^{D(3)}=\frac{16}{(2\pi)^3}\frac{Q^6}{\beta}N_{c}\int_0^1
dz\, (z (1-z))^3 } \nonumber \\
&&\hspace*{-2em} \times \int_0^\infty db\, b \int_0^{2\pi} d\theta  \nonumber \\
&&\hspace*{-2em} \times \left| \int \frac{d^2 r_1}{4\pi}\; e^{\vec{l}\cdot\vec{r}_1} {\rm K}_0(\epsilon r_1)W\left(\vec{r}_1,b,z\right)\right|^2 \; .
\end{eqnarray}

\item[b)]Transversal case:
\begin{eqnarray} \label{transversal}
\lefteqn{x_P F_{\rm T}^{D(3)}=\frac{4}{(2\pi)^3}\frac{Q^6}{\beta}N_{c}\int_0^1 dz\, (z (1-z))^2} \nonumber \\
&&\hspace*{-2em} \times \; (z^2+ (1-z)^2) \int_0^\infty db\, b \int_0^{2\pi} d\theta \nonumber  \\ 
&&\hspace*{-2em} \times \left| \int \frac{d^2 r_1}{4\pi}\; e^{i\theta_{r_1}}\; e^{\vec{l}\cdot\vec{r}_1}{\rm K}_1(\epsilon r_1) W\left(\vec{r}_1,b,z\right)\right|^2 \; .
\end{eqnarray}
\end{description}

\noindent where \mbox{$\vec{l}= \epsilon \sqrt{\frac{1-\beta}{\beta}}\left(\begin{array}{c}\cos\theta\\\sin \theta\end{array}\right)$ with $\epsilon=\sqrt{z(1-z)Q^2}$}, \\ is the transverse momentum of the $q-\bar{q}$ pair.

These results have the same structure as those of Buchm\"uller et al. \cite{BuchmullerMcDemontHebecker}, but we go further and calculate the dipole-proton scattering in the MSV and give numerical results.

\section{Results}

In this section we present the numerical evaluation of equations (\ref{longitudinal}) and (\ref{transversal}). The error bars in the plots come from the numerical computation. We have taken the parameters of the model, \[a=0.31fm,\,<g^2FF>=3GeV^4,\,S_P=0.85fm \] as in \cite{RueterPhD}.
The structure functions are calculated in the range from $1.0$ to $12.0$ $GeV^2$ for $Q^2$. For the $\beta$ dependence we take values between $0.5$ and $1.0$ where the non-perturbative contribution becomes more important because the transverse momentum of the quarks gets small making that the eikonal factors ``feel'' large transverse distances. Since the extreme regions on the $z$ integral and small dipole radius might give an important contribution we present also results using the modifications to the model proposed by R\"uter \cite{Rueterthis}. 

\begin{figure}[h]
\leavevmode 
\begin{center}
\begin{picture}(200,125)
\put(0,70){\rotatebox{90}{$x_{\rm p} F_{\rm T}^{D(3)}$}}
\put(10,0){\includegraphics[width=6.5cm]{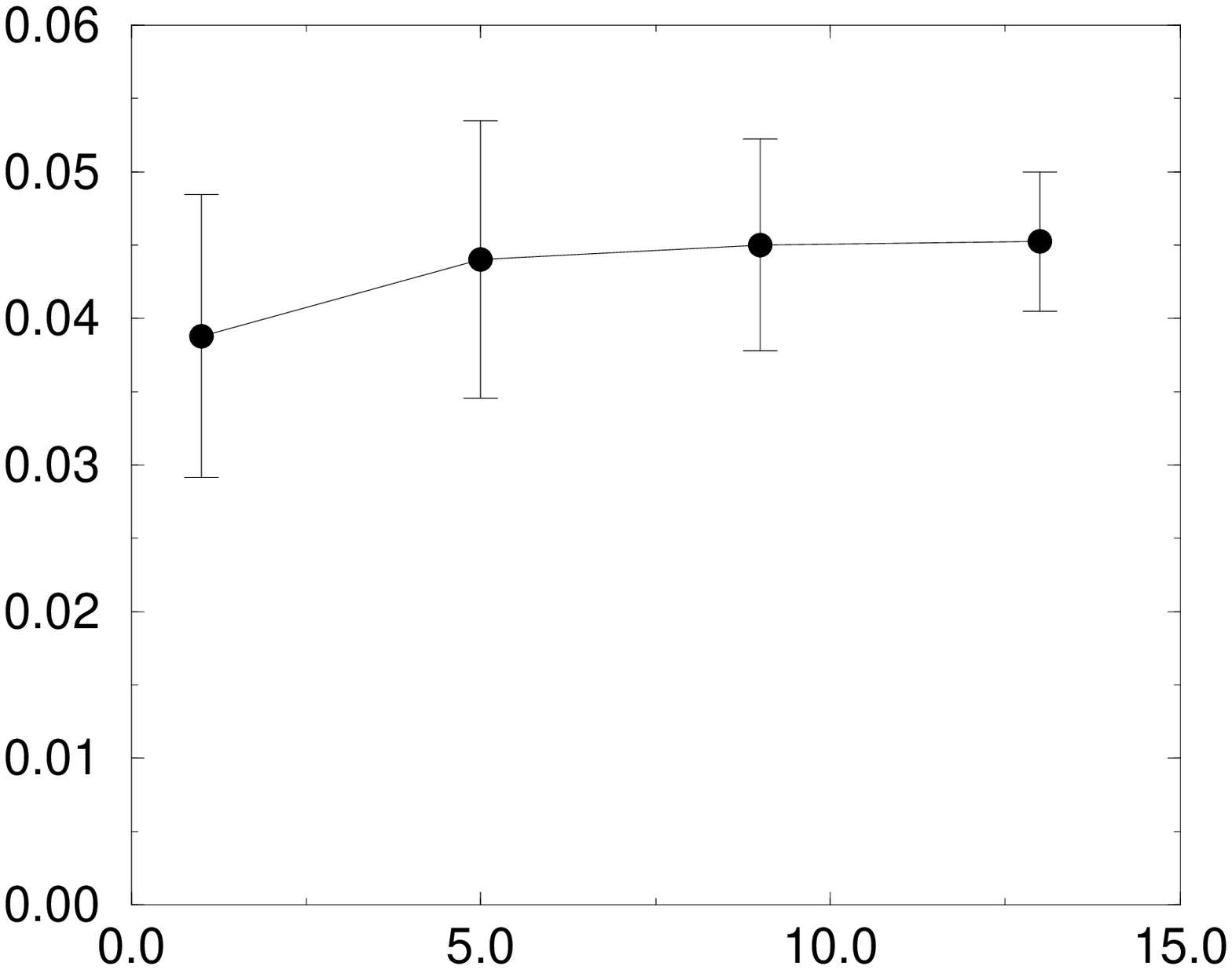}}
\put(90,0){$Q^2 \; \; GeV^2$}
\end{picture}
\unitlength1cm
\vspace*{-2em}
{\caption{\scriptsize Transversal structure function for fixed $\beta = 0.65$ as a function of $Q^2$ (In this and in the following plots the error bars shown come from the numerical computation).}\label{tb65qcd}}
\end{center}
\end{figure}

For the behavior of the structure functions against the photon's momentum $Q^2$ at fixed $\beta = 0.65$ we found out that the transversal structure function, as shown in figure \ref{tb65qcd}, gets constant at big values of $Q^2$ revealing a leading twist behavior. The same $Q^2$ independence can be seen for different values of $\beta$.
 
\vspace*{-2em}

\begin{figure}[h]
\leavevmode 
\begin{center}
\begin{picture}(200,125)
\put(0,70){\rotatebox{90}{$x_{\rm p} F_{\rm L}^{D(3)}$}}
\put(10,0){\includegraphics[width=6.5cm]{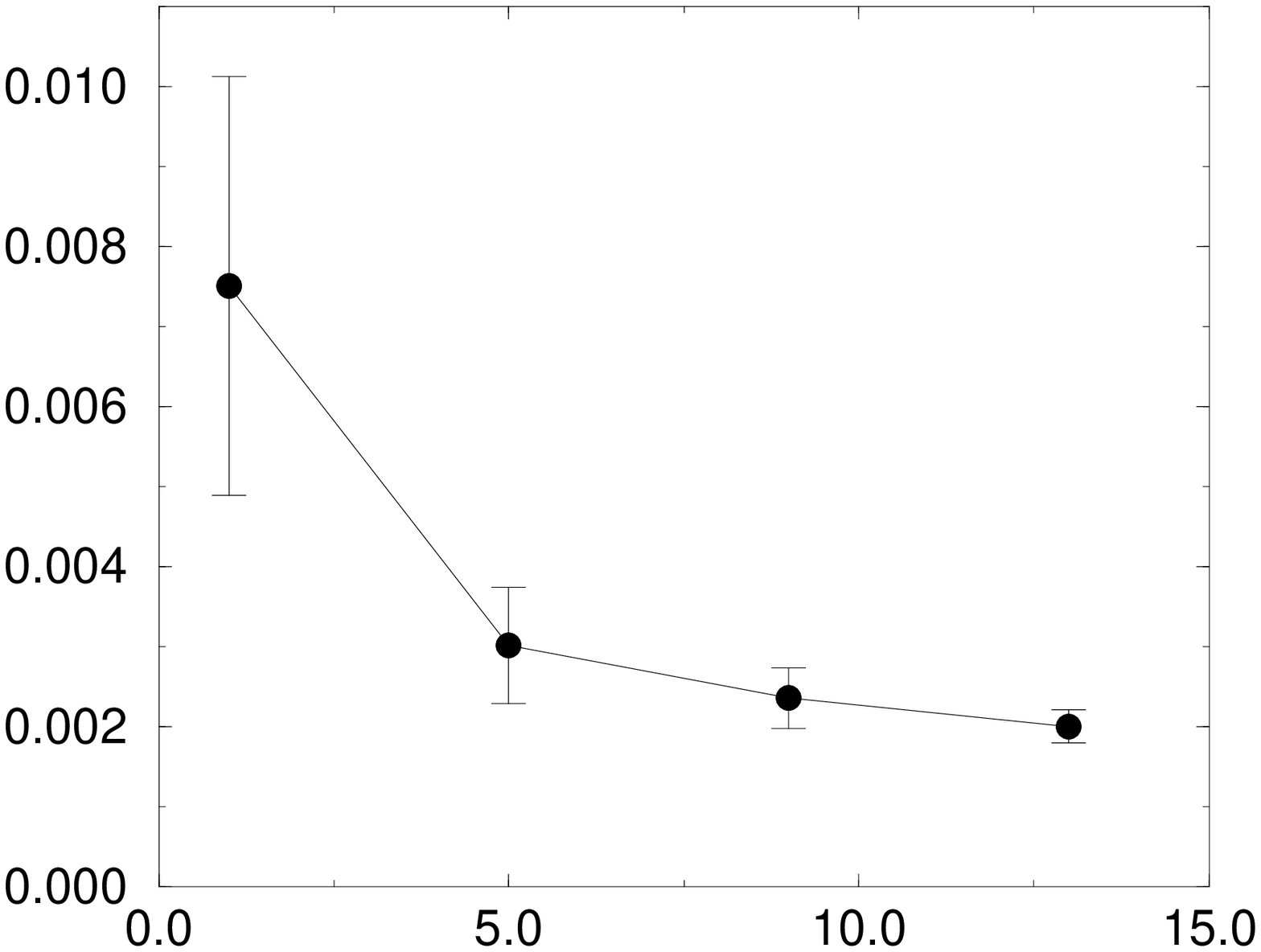}}
\put(90,0){$Q^2 \; \; GeV^2$}
\end{picture}
\unitlength1cm
\vspace*{-2em}
{\caption{\scriptsize Longitudinal structure function for fixed $\beta = 0.65$ as a function of $Q^2$.}\label{lb65qcd}}
\end{center}
\end{figure}

The longitudinal structure function (figure \ref{lb65qcd}) as a function of $Q^2$ for fixed $\beta=0.65$ has an important contribution at low values of $Q^2$ and decreases with increasing $Q^2$ showing a higher twist behavior. The same behavior is observed for di\-ffer\-ent values of $\beta$.

\begin{figure}[h]
\leavevmode 
\begin{center}
\begin{picture}(200,125)
\put(10,0){\includegraphics[width=6.5cm]{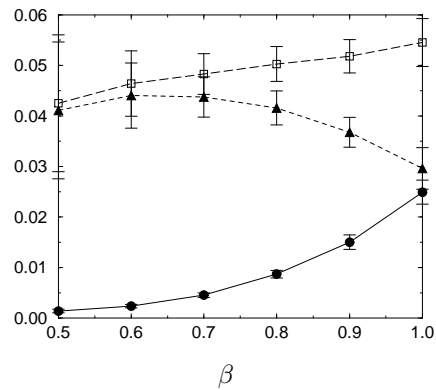}}
\put(100,0){$\beta$}
\end{picture}
\unitlength1cm
\vspace*{-2em}
{\caption{\scriptsize $x_{\rm p} F_{\rm L}^{D(3)}$ (solid), $x_{\rm p} F_{\rm T}^{D(3)}$ (dashed) and $x_{\rm p} F_{\rm 2}^{D(3)}$ (long dashed) for fixed $Q^2=4.5 \; GeV^2$ as a function of $\beta$.}\label{figura4}}
\end{center}
\end{figure}

\begin{figure}[h]
\leavevmode 
\begin{center}
\begin{picture}(200,125)
\put(10,0){\includegraphics[width=6.5cm]{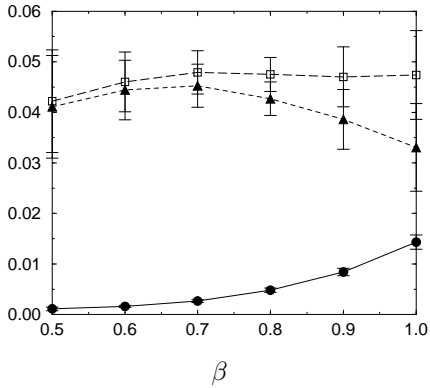}}
\put(100,0){$\beta$}
\end{picture}
\unitlength1cm
\vspace*{-2em}
{\caption{\scriptsize $x_{\rm p} F_{\rm L}^{D(3)}$ (solid), $x_{\rm p} F_{\rm T}^{D(3)}$ (dashed) and $x_{\rm p} F_{\rm 2}^{D(3)}$ (long dashed) for fixed $Q^2=12.0 \; GeV^2$ as a function of $\beta$.}\label{figura6}}
\end{center}
\end{figure}

The $\beta$ dependence of the structure functions at fixed values of $Q^2$ is plotted in figure \ref{figura4} for $Q^2=4.5 \; GeV^2$ and in figure \ref{figura6} for $Q^2=12.0 \; GeV^2$. The longitudinal structure function grows up with $\beta$ whereas the transversal decreases. As $\beta \rightarrow 1$ the longitudinal contribution to $x_{\rm p} F_{\rm 2}^{D(3)}$ becomes important. These results are not changed by the modifications introduced in the model.

\vspace*{-3em}

\vspace*{-2em}

\begin{figure}[h]
\leavevmode 
\begin{center}
\begin{picture}(200,125)
\put(10,0){\includegraphics[width=6.5cm]{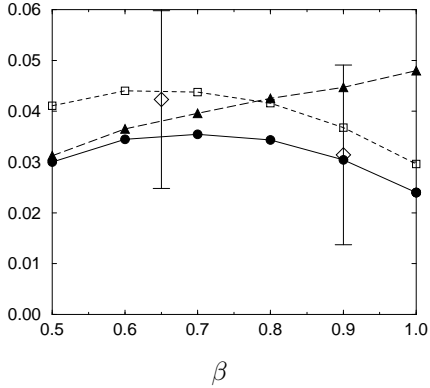}}
\put(100,0){$\beta$}
\end{picture}
\unitlength1cm
\vspace*{-2em}
{\caption{\scriptsize $x_{\rm p} F_{\rm T}^{D(3)}$ in the non modified model (dashed), $x_{\rm p} F_{\rm 2}^{D(3)}$ (long dashed) and $x_{\rm p} F_{\rm T}^{D(3)}$ (solid) with modifications in the model for fixed $Q^2=4.5 \; GeV^2$ \protect\cite{Rueterthis} as a function of $\beta$ compared with experimental points (diamonds).}
\label{figura8}}
\end{center}
\end{figure}

\begin{figure}[h]
\leavevmode 
\begin{center}
\begin{picture}(200,125)
\put(10,0){\includegraphics[width=6.5cm]{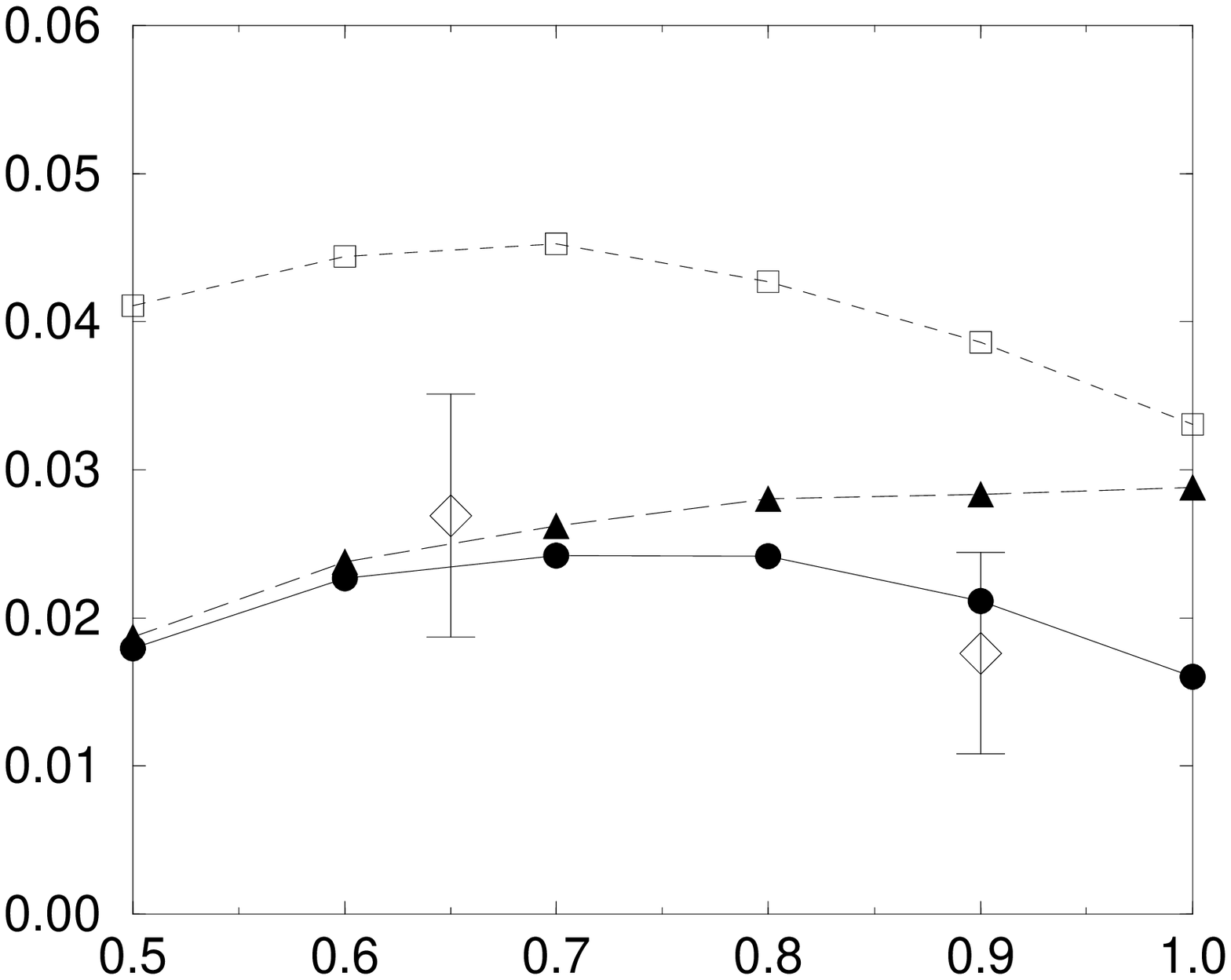}}
\put(100,0){$\beta$}
\end{picture}
\unitlength1cm
\vspace*{-2em}
{\caption{\scriptsize $x_{\rm p} F_{\rm 2}^{D(3)}$ in the non modified model (dashed), $x_{\rm p} F_{\rm 2}^{D(3)}$ (long dashed) and $x_{\rm p} F_{\rm T}^{D(3)}$ (solid) with modifications in the model for fixed $Q^2=12.0 \; GeV^2$ \protect\cite{Rueterthis} as a function of $\beta$ compared with  experimental points (diamonds).}\label{figura9}}
\end{center}
\end{figure}

We also compared our predictions for $x_{\rm p} F_{\rm 2}^{D(3)}$ and $x_{\rm p} F_{\rm T}^{D(3)}$ with the experiment \cite{HERAH1}. We see that at $Q^2=4.5 \; GeV^2$ (figure \ref{figura8}) we have a good agreement with experiment, but for $Q^2=12.0 \; GeV^2$ (figure \ref{figura9}) the result of the no modified model is by a factor 2 to large. Introducing modifications for small dipoles as proposed in ref \cite{Rueterthis} leads however to very good agreement.

\vspace*{-2em}

\vspace*{-3em}

\section{Conclusions}
Our approach for calculating the diffractive structure functions is similar to the semi-classical approach of Buchm\"uller et al. \cite{BuchmullerMcDemontHebecker}. Our calculation reproduces the leading twist behavior for $x_{\rm p} F_{\rm T}^{D(3)}$, but gives a not negligible contribution at $\beta$ near $1$. It was possible to get good agreement with experiment introducing the modifications presented in \cite{Rueterthis}. Further work to calculate the structure functions for values of $Q^2 < 1.0 \; GeV^2$ is in progress.

\vspace*{-2em}

\section{Acknowledgments}
I would like to thank H.G. Dosch, A. Hebecker, M. R\"uter, G. Kulzinger and E. Berger for useful discutions and comments.


\begin{thebibliography}{99}
\bibitem{Nachtmann} O. Nachtmann, Ann. Phys. {\bf 209}, 436 (1990). 
\bibitem{Dosch1} H.G. Dosch, Phys. Lett. {\bf B190}, 177 (1987).
\bibitem{DoschSimonov} H.G. Dosch and Y.A. Simonov, Phys. Lett. {\bf B205}, 339 (1988).
\bibitem{BuchmullerMcDemontHebecker}W. Buchm\"uller, M.F. McDermott and A. Hebecker, Nucl. Phys. {\bf B487} (1997) 283.
\bibitem{DoschFerreraKramer} H.G. Dosch, E. Ferreira, A. Kr\"amer, Phys. Rev. {\bf D50}, 1992 (1994).
\bibitem{DoschGousKulPir} H.G. Dosch, T. Gousset, G. Kulzinger and H.J. Pirner, Phys. Rev. {\bf D55}, 2602 (1997).
\bibitem{RueterPhD} M. R\"uter, {\it Quark-Confinement und diffraktive Hadron-Streuung im Modell des stochastischen Vakuums}, Ph.D. Thesis (Univ. Heidelberg), 1997. 
\bibitem{Rueterthis}M. R\"uter, These proceedings.
\bibitem{HERAH1} C. Adloff et al. Z. Phys. {\bf C76} (1997) 613.
%

\end{thebibliography}
\end{document}